\documentclass[aps,prl,reprint,superscriptaddress]{revtex4-2}
\usepackage{amsmath,amssymb}
\usepackage{graphicx}
\usepackage[dvipsnames]{xcolor}

\usepackage{physics}
\usepackage{hyperref}
\usepackage{bm}

\begin{document}

\title{
The cost of speed: Time-optimal thermal control of trapped Brownian particles
}

\author{Miguel Ib\'a\~nez}
\affiliation{Universidad de Granada, Department of Applied Physics and Research Unit ‘Modeling Nature’ (MNat), Granada, Spain}
\affiliation{Nanoparticle Trapping Laboratory, Universidad de Granada, Granada, Spain}
                        
\author{Antonio Patr\'on-Castro}
\affiliation{Department of Physics, Simon Fraser University, Burnaby, British Columbia V5A 1S6, Canada}

\author{Antonio Lasanta}
\affiliation{Universidad de Granada, Department of Algebra, Facultad de Educación, Economía y Tecnología de Ceuta, Ceuta, Spain}
\affiliation{Instituto Carlos I de F\'isica Te\'orica y Computacional, Universidad de Granada, E-18071 Granada, Spain}
\affiliation{Nanoparticle Trapping Laboratory, Universidad de Granada, Granada, Spain}

\author{Carlos A.~Plata}
\author{A.~Prados}
\affiliation{
  Física Teórica, Multidisciplinary Unit for Energy Science, Universidad de Sevilla, Apartado de Correos 1065, E-41080 Sevilla, Spain
}%

\author{Ra\'ul A. Rica-Alarc\'on}
\email[Corresponding author; email: ]{rul@ugr.es}
\affiliation{Universidad de Granada, Department of Applied Physics and Research Unit ‘Modeling Nature’ (MNat), Granada, Spain}
\affiliation{Nanoparticle Trapping Laboratory, Universidad de Granada, Granada, Spain}

\date{\today}


\begin{abstract}
A thermal analogue of the classical brachistochrone problem, which minimizes the connection time between two equilibrium states of harmonically confined Brownian particles, has recently been solved theoretically. Here we report its experimental realization using two optically trapped microparticles subjected to a bang--bang effective temperature protocol. Despite their distinct relaxation times, both degrees of freedom are steered to their respective equilibrium states simultaneously in a finite minimal time. We provide a complete time-resolved characterization of the nonequilibrium dynamics through the evolution of the position variances and the entropy production within the framework of stochastic thermodynamics, enabling a quantitative comparison with direct relaxation and a suboptimal protocol. In addition, we employ information-geometric tools---recently referred to as thermal kinematics---to track the system's path in state space with a single dynamical quantity. Our results show that faster equilibration requires a larger entropy production and an increased thermodynamic length, revealing a direct trade-off between temporal optimality and thermodynamic cost in multidimensional stochastic systems driven by a single intensive control parameter.
\end{abstract}

\maketitle


\textit{Introduction}---Understanding how swift state-to-state transformations (SSTs) outpace natural relaxation at micro- and nanoscales is a central problem in nonequilibrium physics~\cite{schmiedl_optimal_2007,aurell_optimal_2011,aurell_refined_2012,sivak_thermodynamic_2012,martinez2016engineered,chupeau_engineered_2018,baldassarri_engineered_2020,patron_minimum_2024}---see Ref.~\cite{guery2022} for a recent review. A particularly sharp formulation is the thermal brachistochrone: determining the temperature protocol that connects two equilibrium states at different temperatures in the minimum possible time~\cite{patron2022thermal}, i.e., the direct thermal analogue of the classical brachistochrone---posed by Johann Bernoulli in 1696~\cite{Bernoulli1696_inActa}.

Thermal transformations can be accelerated either under autonomous dynamics---such as in the Mpemba effect and related problems~\cite{lasanta2017,BaityJesi2019,klich2019,kumar2020,Bechhoefer2021,patron_non-equilibrium_2023,patron_strong_2021,ibanez2023,teza2026speedups}---or through externally controlled nonequilibrium protocols, generally accompanied by a thermodynamic cost linked to the irreversibility of the protocol~\cite{chupeau_engineered_2018,plata_finite-time_2020,prados_optimizing_2021,patron2022thermal,pires2023optimal}. For driven transformations, the thermodynamic cost vanishes only in the quasistatic limit, which takes infinite time, thereby highlighting the unavoidable trade-off between speed and dissipation captured by thermodynamic speed limits \cite{Shiraishi2018speed,falasco2020dissipation,Ito2020Stochastic,plata_finite-time_2020,prados_optimizing_2021,dechant_minimum_2022,pires2023optimal,dieball2024thermodynamic,guery2022}.

Most experimental realizations of SST protocols have focused on isothermal and effectively one-dimensional systems~\cite{martinez2016engineered,li_shortcuts_2017,salambo2020engineered,frim2021engineered,baldovin2025optimal,sherf2025stochastic,goerlich2026consistent}. Extending such control to multiple degrees of freedom is challenging, both conceptually and experimentally: it may seem to require as many independent control parameters as dynamical quantities. Within this context, the thermal brachistochrone~\cite{patron2022thermal} demonstrates that a single intensive parameter---the bath temperature---allows simultaneously controlling multiple harmonically confined modes.

Here we report an experimental implementation of thermal brachistochrone protocols in a two-dimensional system subject to experimental constraints. Specifically, we use two Brownian particles trapped by optical tweezers with different stiffnesses, showing that the brachistochrone can drive them between equilibrium states in a finite time, outperforming suboptimal alternatives. Moreover, we provide a complete thermodynamic characterization of these 
nonisothermal protocols through combined theoretical analysis and experimental validation of new features, including their irreversible entropic cost.  

\begin{figure*}
    \centering
    \includegraphics[]{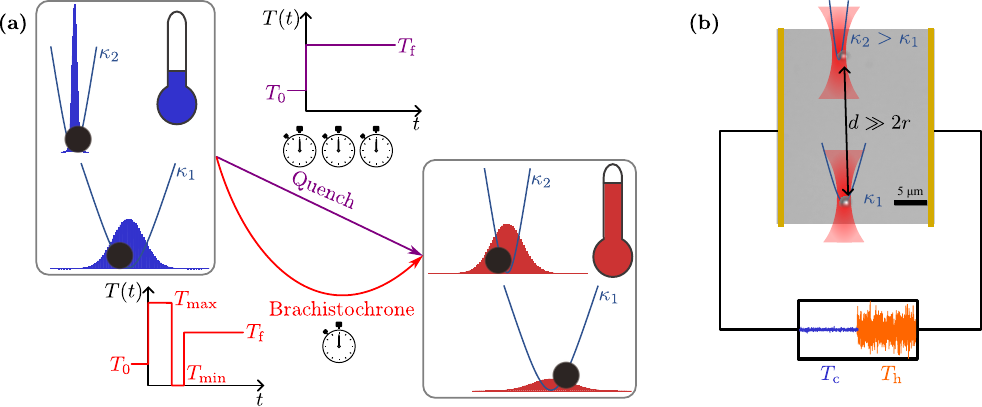}
    \caption{\textrm{Thermal brachistochrone and experimental platform.}
    \textrm{(a)} We connect two equilibrium states by changing only the bath temperature while keeping the trap stiffnesses fixed. A direct quench from $T_0$ to $T_{\rm f}$ (top) relaxes exponentially. The time-optimal thermal brachistochrone (bottom) is a bang-bang protocol that switches between the accessible bounds $T_{\min}$ and $T_{\max}$ and reaches the target state in the minimum possible time for the given values of  $\{T_{\min},T_{\max}\}$.
    \textrm{(b)} Two charged dielectric microspheres dispersed in water are independently confined in harmonic optical traps generated by a split, tightly focused laser beam. The trap centers are separated by a distance $d$ much larger than the particle diameter, so hydrodynamic coupling is negligible. An external, spatially uniform noisy electric field produced by submerged microelectrodes drives the particles and sets an effective bath temperature $T$ above the ambient value, yielding Brownian motion with a Gaussian position distribution in each trap at temperature $T$.
    }
    \label{fig:fig1}
\end{figure*}

\textit{Objective}---We consider a system of $N$ Brownian particles trapped in harmonic potentials with fixed stiffnesses $\kappa_i$ $(i=1,...,N)$, in contact with a thermal bath of controllable temperature $T(t)$---see Fig.~\ref{fig:fig1}(a). The initial and final equilibrium states are specified by bath temperatures $T_0$ and $T_{\rm f}$. Our goal is to connect these states in the shortest possible time by controlling $T(t)$, implementing a thermal brachistochrone.
The time-optimal solution is a bang-bang protocol in which the bath temperature switches between the accessible extrema $T_{\min}$ and $T_{\max}$, with a number of bangs---constant-temperature stages---equal to the number of controlled degrees of freedom~\cite{patron2022thermal}. For heating (cooling), the protocol starts with a quench to $T_{\max}$ ($T_{\min}$), followed by alternating switches until all degrees of freedom simultaneously reach equilibrium at $T_{\mathrm f}$ in a finite time $t_{\rm f}$.


\textit{Experiment and theoretical description}---Our experimental platform consists of $N\!=\!2$ identical, charged silica microspheres of diameter $2r\!=\!1~{\rm \mu m}$ suspended in ultrapure water and manipulated by a commercial dual optical tweezers device---{see Fig.~\ref{fig:fig1}(b)}. Two independent traps are generated from a single $1064~{\rm nm}$ laser beam that is split into two paths, each independently steered and tightly focused to a diffraction-limited spot by a high numerical aperture objective inside a fluid chamber, thereby producing a nearly harmonic trapping potential at each focus. In our experiments, the two trap centers are separated by $d\sim 20~{\rm \mu m}$, so hydrodynamic coupling can be neglected and the particles treated as dynamically independent \cite{salambo2020engineered}. The trap stiffnesses $\kappa_i$ are controlled by the optical power at each focus and remain fixed during each run; we characterize the system by the stiffness ratio $\kappa_2/\kappa_1$. Particle positions in the XY plane are measured with a quadrant photodiode (QPD), whose differential voltage is proportional to the displacement from the corresponding trap center. 

Inertia can be safely neglected, since the momentum relaxation time for silica in water is about 0.1~$\mu$s and the trap relaxation time of particles in the trap is $\sim$1~ms in our experimental conditions---see below. Therefore, we describe the motion for particle $i$ with the overdamped Langevin equation, which reads
\begin{equation}
    \gamma \dot{x}_i(t) = -\kappa_i x_i(t) + \xi_i(t),
\end{equation}
where $x_i(t)$ is its displacement from the trap center, $\gamma=6\pi\eta r$ is the Stokes friction coefficient, and $\xi_i(t)$ is a Gaussian white noise, $\langle \xi_i(t)\rangle = 0$ and $\langle \xi_i(t)\xi_j(t')\rangle = 2\gamma k_{\rm B}T(t)\,\delta_{ij}\delta(t-t')$.
Starting from an equilibrium state, the joint process $\{x_1,x_2\}$ is Gaussian with zero mean and correlation, $\expval{x_i(t)}=0$ and $\expval{x_1(t) x_2(t)}=0$, for all times. Therefore, it is completely characterized by the variances $s_i(t)\equiv \langle x_i^2(t)\rangle$, which evolve according to
\begin{equation}\label{eq:variance}
    \gamma \dot{s}_i(t) = -2\kappa_i s_i(t) + 2 k_{\rm B}T(t).
\end{equation}
Equation~\eqref{eq:variance} identifies the natural relaxation time $\tau_i\equiv \gamma/\kappa_i$. At equilibrium, the Gibbs distribution is recovered with variance $s_i^{\rm eq}=k_{\rm B}T/\kappa_i$, matching the steady-state solution of Eq.~\eqref{eq:variance}.

Nonisothermal protocols are implemented by taking advantage of the charge that silica particles acquire when immersed in water. A noisy, spatially uniform, electric field directed along the $x$ axis is generated by a pair of gold plated tungsten wires (diameter 50~$\mu$m) arranged as parallel electrodes in the fluid chamber. The applied voltage is a Gaussian white-noise signal with time-dependent amplitude, piecewise constant in our implementation, which effectively increases the bath temperature without modifying viscous dissipation \cite{martinez2013effective}. The dynamics is therefore equivalent to those of a system of Brownian particles coupled to an effective bath at temperature $T(t)=T_{\rm b}+\Delta T(t)$, where $T_{\rm b}=293~\mathrm{K}$ is the ambient temperature and $\Delta T(t)\propto s_V(t)\ge 0$ is controlled by the instantaneous variance $s_V$ of the electrical noise. Importantly, we remark that every particle will experience a different $T$ due to the fact that each particle has a different charge, but this does not prevent implementing the brachistochrone protocol with a single control parameter. The mapping between $s_V$ and $\Delta T$ is discussed in~\cite{SM} and, in what follows, we refer to the effective temperature $T$ as the bath temperature for clarity.

In practice, the temperature cannot be tuned arbitrarily. In our experiments, the minimum attainable value coincides with the ambient bath temperature, $T_{\min}=T_{\rm b}$, i.e., that of the water where experiments are performed. The upper bound is set by the maximum output of the signal generator, $T_{\max}=T_{\rm b}+(\Delta T)_{\max}$, with $(\Delta T)_{\max}\propto s_{V,\max}$. 
Let us consider a finite-time connection, with total duration $t_{\rm f}$, from an initial equilibrium state at temperature $T_0$ to a final equilibrium state at temperature $T_{\rm f}$~\footnote{For all connecting protocols $T(t)$, the temperature of the bath is instantaneously changed to $T_{\rm f}$ at $t_{\rm f}^+$ to keep the system in equilibrium at the final temperature for $t>t_{\rm f}$.}. Then, we have the boundary conditions
\begin{equation} \label{eq:bc}
s_i(0)=s_{i,0}\equiv k_{\rm B}T_0/\kappa_i,\quad
s_i(t_{\rm f})=s_{i,\rm f}\equiv k_{\rm B}T_{\rm f}/\kappa_i.
\end{equation}
During each stage of a bang-bang protocol, the temperature is constant and equal to either $T_{\min}$ or $T_{\max}$ for a time span $\Delta t_i$, in which $s_i(t)$ relaxes exponentially. By concatenating $N$ such exponential evolutions with the boundary conditions~\eqref{eq:bc}, a closed set of $N$ equations for the time durations $\Delta t_i$ is obtained. The total connection time is then $t_{\rm f}=\sum_{i=1}^{N}\Delta t_i$, which is the minimum connection time between the initial and target states---see Ref.~\cite{patron2022thermal} and \cite{SM} for the equations defining $\Delta t_i$.

\begin{figure}
    \centering
    \includegraphics[width=0.4\textwidth]{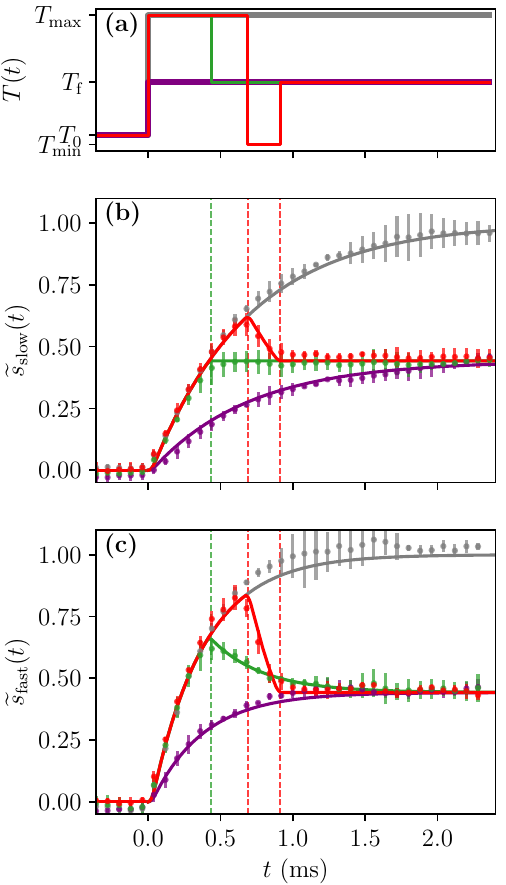}
    \caption{\textrm{Two-particle variance dynamics under heating protocols connecting the same equilibrium states.}
    \textrm{(a)} Bath temperature $T(t)$ for the three heating protocols applied to the same two-particle system: direct relaxation (purple), suboptimal one-bang protocol (green), and optimal two-bang (brachistochrone) protocol (red).
    \textrm{(b)} and \textrm{(c)} Corresponding rescaled variances
    $\widetilde{s}_i(t)$ for the slow (with a characteristic time $\tau_{\rm slow} = 1.35 \ {\rm ms}$) and fast ($\tau_{\rm fast} = 0.72 \ {\rm ms}$) particles. Vertical dashed lines mark the instants at which the temperature jumps in each protocol, with the same color code. Symbols are experimental data and solid lines are the theoretical predictions without fitting parameters. Error bars denote the standard error of the mean over 11400 repeated runs of each protocol.
    Gray curves correspond to the direct relaxation $T_0\to T_{\max}$ shown as a reference.}
    \label{fig:fig2}
\end{figure}

\textit{Thermal brachistochrone of two trapped particles}---Although the above theoretical analysis applies for arbitrary $N$, here we focus on systems with two degrees of freedom. 
The two Brownian particles have distinct relaxations times $\tau_i$, with $\tau_1>\tau_2$: We thus refer to particle~1 and~2 as the slow and fast one, respectively.
In this case, the thermal brachistochrone comprises two bangs alternating between $T_{\min}$ and $T_{\max}$.

Figure~\ref{fig:fig2} presents the experimental time evolution of the variances during a heating brachistochrone, together with their 
relaxations when directly quenched to the target temperature $T_{\rm f}$. It also includes the evolution under a \textit{suboptimal protocol}, defined as the optimal one-dimensional thermal brachistochrone for the slowest particle---see \cite{SM} for an extended analysis of the one-dimensional case. Specifically, we show the rescaled variances $\widetilde{s}_i(t)\equiv [s_i(t)-s_i(0)]/|s_{i,\max}-s_i(0)|$, where $s_{i,\max}=k_{\rm B}T_{\max}/\kappa_i$.

In the direct relaxation at constant temperature $T_{\rm f}$, each variance relaxes from its initial value $s_{i,0}$ to its target value $s_{i,\rm f}$, reaching the latter only after infinite time. In the suboptimal protocol, the system relaxes at the maximum temperature $T_{\max}$ until the slow particle 1 reaches equilibrium at $T_{\rm f}$ at $t=\Delta t$, $s_{1}(\Delta t)=s_{1,\rm f}$, and $T(t)=T_{\rm f}$ for $t>\Delta t$. However, the fast particle 2 has overshot its equilibrium $s_{2,\rm f}$ at $t=\Delta t$, reaching it---again---only after infinite time. 

In contrast, the brachistochrone achieves the same connection in a finite  time. Specifically, it comprises two bangs: (i) the first one with $T(t)=T_{\max}$ with duration $\Delta t_1$, i.e., $0<t<\Delta t_1$, (ii) the second one with $T(t)=T_{\min}$ with duration $\Delta t_2$, i.e., $\Delta t_1<t<t_{\rm f}=\Delta t_1+\Delta t_2$. This two-bang protocol induces a transient overshoot of both variances, which is necessary for the synchronized arrival of the slow and fast particle at the target equilibrium. During the first bang, $0\!<\!t\!<\!\Delta t_1$, both particles follow the same evolution as under a quench to $T_{\max}$, overshooting their final values; the subsequent bang, $\Delta t_1\!<\!t\!<\!t_{\rm f}=\Delta t_1+\Delta t_2$, reverses the dynamics, bringing both variances exactly to their target at $t=t_{\rm f}$. Although this transient overshoot may appear reminiscent of a Kovacs-like hump~\cite{kovacs_isobaric_1979,prados_kovacs_2010,patron_strong_2021,Prados2014,militaru2021}, its origin is fundamentally different. Here, each degree of freedom is directly resolved and controlled, and the overshoot is induced by the time-optimal protocol rather than arising from hidden internal relaxation or memory effects.

The brachistochrone outperforms both the direct relaxation at $T_{\rm f}$ and the suboptimal protocol. While the suboptimal protocol brings the slow particle to the target slightly earlier, the optimal brachistochrone reaches the final equilibrium of both degrees of freedom simultaneously in a finite time. This behavior is consistent with the theoretical prediction that, for multiple degrees of freedom, the minimal connection time depends nontrivially on the relaxation-time spectrum and features a singular degenerate limit \cite{patron2022thermal}; in particular, applying the optimal, one-dimensional protocol to the two-particle system generally leaves an overshoot of the fast mode that must be corrected, revealing a structural gap between the optimal control strategies in one and two-particle systems.

\textit{Thermal kinematics}---While the variance traces in Fig.~\ref{fig:fig2} provide a degree-of-freedom–resolved description of the dynamics, a multidimensional transformation can be more compactly characterized by a single scalar quantity. Thermal kinematics provides such a description by endowing the space of probability distributions with a geometric structure \cite{Ito2020Stochastic,li_geodesic_2022,li_geodesic_2023,ibanez2023}.

Within this framework, the evolution can be characterized by the thermodynamic length $\mathcal{L}(t)$, defined in terms of the Fisher information $I(t)$ associated with the time-dependent distribution $p(\bm{x},t)$, $\bm{x}\equiv\{x_1,x_2\}$,
\begin{equation}
\mathcal{L}(t)\equiv \int_{0}^{t}\mathrm{d}t'\sqrt{I(t')}, \quad 
I(t)
\equiv \!\int \!\mathrm{d}\bm{x}\,
p(\bm{x},t)\left[\partial_t \ln p(\bm{x},t)\right]^2 .
\end{equation}
where $\sqrt{I(t)}$ defines an intrinsic speed in distribution space \cite{Ito2020Stochastic}. We further introduce the normalized degree of completion $\varphi(t)\equiv \mathcal{L}(t)/\mathcal{L}_{\rm tot}$ with respect to the thermodynamic length of the complete trajectory~\footnote{Note that $\mathcal{L}_{\rm tot}=\mathcal{L}(\infty)$ in the direct and suboptimal protocols, while $\mathcal{L}_{\rm tot} = \mathcal{L}(t_{\rm f})$ in the optimal protocol.}, which increases monotonically from zero to unity~\cite{ibanez2023}.

\begin{figure}
    \centering
    \includegraphics[width=0.4\textwidth]{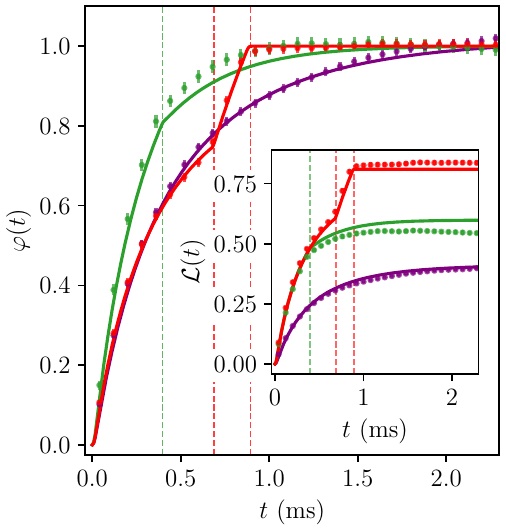}
    \caption{{Thermal-kinematics characterization of the two-particle heating protocols.}
    Degree of completion $\varphi(t)=\mathcal{L}(t)/\mathcal{L}_{\rm tot}$ for the full two-particle system under direct relaxation (purple), the suboptimal protocol (green), and the optimal (brachistochrone) protocol (red), for the same data set as Fig.~\ref{fig:fig2}. Again, symbols are experimental data and solid lines are the corresponding theoretical predictions with no fitting parameters. Vertical dashed lines mark the instants at which the temperature jumps during the implementation of each protocol, with the same color code. Error bars are obtained by quadratic propagation of the experimental uncertainties of the variances.
    \textit{Inset}: Accumulated thermodynamic length $\mathcal{L}(t)$ as a function of time for the same protocols and color code, showing that the optimal protocol sweeps a longer distance in distribution space in a shorter time.}
    \label{fig:fig3}
\end{figure}

Figure~\ref{fig:fig3} shows $\varphi(t)$ and $\mathcal{L}(t)$ for the direct, optimal, and suboptimal protocols in Fig.~\ref{fig:fig2}. While both optimal and suboptimal protocols initially progress similarly---see the inset in Fig.~\ref{fig:fig3}, the brachistochrone reaches $\varphi=1$ in a finite time. Moreover, different protocols reaching the same final state can traverse markedly different thermodynamic lengths: the time-optimal protocol covers a significantly larger distance in distribution space over a shorter time, corresponding to a higher speed. Thermal kinematics thus exposes a trade-off between connection time and the extent of nonequilibrium exploration, providing a compact and experimentally accessible benchmark for multidimensional stochastic control.

\textit{Energetic and entropic cost}---Connection time alone does not quantify the operational cost of a transformation: in general, there are many protocols sharing the same connection time with different degrees of irreversibility. Within the framework of stochastic thermodynamics, protocol-dependent quantities like work and heat are usually employed to quantify irreversibility~\cite{schmiedl_optimal_2007,aurell_optimal_2011,aurell_refined_2012,sivak_thermodynamic_2012,martinez2016engineered,guery2022}. However, in the isochoric nonisothermal connections considered here, the mechanical work vanishes identically and the ensemble-averaged heat is fixed by the endpoint equilibrium states. Therefore, while these energetic quantities fail to distinguish between protocols connecting the same equilibrium states, the entropy production---being genuinely path-dependent---provides a direct measure of the dissipation generated along the nonisothermal transformation~\cite{pires2023optimal,rademacher2022nonequilibrium}.

For an overdamped harmonic particle, the total entropy production of particle $i$ reads at the trajectory level $\delta\sigma_i=(k_{\rm B}T-\kappa_i x_i^2) \ {\rm d}T/(2T^2)$, whose average along a protocol $T(t)$ yields
\begin{equation}
\Sigma_i(t)=\frac{k_{\rm B}}{2}\ln \frac{T(t)}{T_0} + \frac{k_{\rm B}\,\kappa_i}{2}\int_0^t {\rm d}t'\ \dot{\beta}(t')\  s_i(t') \ge 0,
\end{equation}
where $\beta^{-1}(t)=k_{\rm B}T(t)$. Summing over $i$, the first term gives the entropy change between the equilibrium states corresponding to the initial temperature $T_0$ and the instantaneous temperature $T(t)$ at time $t$, whereas the second term is the path-dependent entropy production due to irreversibility~\footnote{In some papers, the term \emph{thermal} (or \emph{entropic}) work has been used~\cite{rademacher2022nonequilibrium}.}.  $\Sigma_i$ can be interpreted as the entropic cost of the irreversible process for the $i$-th particle, and $\Sigma(t)=\Sigma_1(t)+\Sigma_2(t)$ as the entropic cost of the whole system. Note that, for discrete temperature jumps, $\Sigma_i(t)$ is piecewise constant in time---see \cite{SM}. 

\begin{figure}[hbtp]
    \centering
    \includegraphics[width=0.45\textwidth]{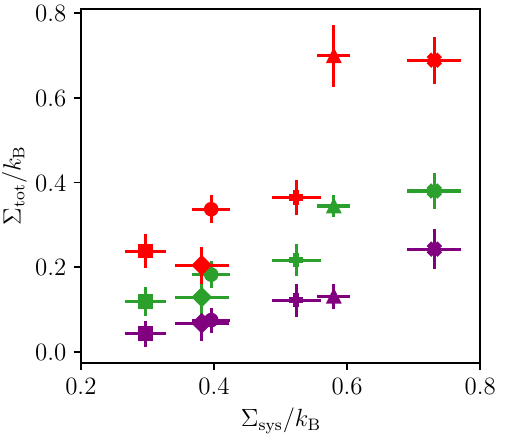}
    \caption{\textrm{Entropic cost for two-particle heating protocols.}
    Total entropy production $\Sigma_{\rm tot}$ 
    as a function of the system entropy change for the direct protocol (purple), the suboptimal protocol (green), and the optimal (brachistochrone) protocol (red). Each symbol corresponds to a distinct experimental series (different control bounds, endpoint temperatures, and stiffness ratios), while data for the same $\Sigma_{\rm sys}$ correspond to the different protocols for the same initial and final states. The hierarchy given by Eq.~\ref{eq:inequality} is always fulfilled. 
    }
    \label{fig:fig4}
\end{figure}

Figure~\ref{fig:fig4} compares the total entropy production of the two-particle system, $\Sigma_{\rm tot}\equiv\Sigma(t_{\rm f})$, with the entropy change between the initial and final states, $\Sigma_{\rm sys}=k_{\rm B}\ln \left(T_{\rm f}/T_0\right)$, 
for direct, suboptimal, and optimal protocols. 
The hierarchy
\begin{equation}\label{eq:inequality}
\Sigma_{\rm tot}^{\rm (opt)}>\Sigma_{\rm tot}^{\rm (sub)}>\Sigma_{\rm tot}^{\rm (dir)},
\end{equation}
which holds generally and admits a rigorous mathematical proof~\cite{SM}, is clearly seen in all the different experimental conditions we explored---see also \cite{SM}, where 
we provide a thorough analysis of the total entropy distributions for each of the three classes of protocols considered here. Equation~\eqref{eq:inequality} captures the principle that outperforming natural relaxation dynamics requires stronger nonequilibrium driving and hence entails greater irreversibility. 

\textit{Discussion}---We have experimentally realized a thermal brachistochrone---the time-optimal temperature protocol connecting equilibrium states of Brownian harmonic oscillators---in a two dimensional setting. 
For two trapped particles with different stiffnesses and thus different relaxation times, the predicted bang-bang protocol enforces a genuinely multidimensional nonequilibrium evolution: a transient overshoot in both degrees of freedom is required for the simultaneously arrival at the target equilibrium state in a finite time. 

The measured variance dynamics agrees quantitatively with the theory with no fitting parameters, demonstrating that control of a single intensive parameter--the bath temperature--suffices to steer multiple independent particles. Analogous cooling experiments display the same behavior and are fully captured by the same theoretical framework, see~\cite{SM} for a full parallel experimental study. 

Beyond coordinate-resolved observables, we have employed recent tools from thermal kinematics to characterize the evolution through a single scalar variable. The thermodynamic length provides a compact comparison between protocols, revealing that the brachistochrone drives the system faster through probability-distribution space while traversing a longer distance. From a thermodynamic standpoint, isochoric nonisothermal driving requires path-sensitive metrics: the total entropy production thus offers a natural operational cost. We derived a robust hierarchy for entropy production, Eq.~\eqref{eq:inequality}, confirming experimentally that accelerating relaxation necessarily entails increased irreversibility.

Our results establish thermal brachistochrones as a minimal-control alternative to many shortcut protocols that rely on multiple knobs or multiobjective optimization. Recent theoretical proposals for the irreversible counterparts of the classic Carnot and Stirling heat engines~\cite{plata_building_2020,prieto-rodriguez_maximum-power_2025} have employed minimum-time branches to maximize the output power. Interestingly, the efficiency at maximum power approaches~\cite{plata_building_2020}---or in some cases exceeds~\cite{prieto-rodriguez_maximum-power_2025}---the Curzon-Ahlborn value~\cite{curzon_efficiency_1975}. The experimental implementation of these brachistochrone branches would thus help 
heat engines improve their power at fixed efficiency toward saturating power-efficiency tradeoffs \cite{blickle2012realization,martinez2016Brownian,pietzonka2018universal,krishnamurthy2023overcoming}.

\begin{acknowledgments}
\textit{Acknowledgments}---This work was supported by MICIU/AEI/10.13039/501100011033 and FEDER, UE under Grants PID2024-161166NB-I00 and PID2024-155268NB-I00;  by C-EXP-251-UGR23 co-funded by Consejería de Universidad,
Investigación e Innovación and by ERDF Andalusia Program 2021-2027; and by the applied research and innovation project PPIT2024-31833 cofunded by EU–Ministerio de Hacienda y Función Pública–Fondos Europeos–Junta de Andalucía–Consejería de Universidad, Investigación e Innovación.
M.I. acknowledges support from an FPU fellowship (FPU21/02569) granted by Ministerio de Ciencia, Innovaci\'on y Universidades (Spain). A.P.C. acknowledges support from the Natural Sciences and Engineering Research Council of Canada (NSERC) Discovery Grant and Discovery Accelerator Supplement RGPIN-2020-04950.
\end{acknowledgments}

\bibliography{references}

\end{document}